\documentclass[11pt]{article}
\usepackage{moriond,epsfig}

\def\Journal#1#2#3#4{{#1} {\bf #2}, #3 (#4)}

\def\PLB{{\em Phys. Lett.}  B}

\def\ZPC{{\em Z. Phys.} C}
\def\JINST{{\em JINST}}
\def\JHEP{{\em JHEP}}
\def\EPJC{{\em Eur. Phys. J.} C}

\def\ra{\rightarrow}

\def\ko{K^0}

\def\be{\begin{equation}}
\def\ee{\end{equation}}
\def\bea{\begin{eqnarray}}
\def\eea{\end{eqnarray}}

\begin{document}

\vspace*{4cm}
\title{THE CMS EXPERIMENT: STATUS AND FIRST RESULTS}

\author{MARTIJN MULDERS (on behalf of the CMS Collaboration)}

\address{European Organization for Nuclear Research, CERN \\
         CH-1211 Geneva 23, Switzerland}

\maketitle\abstracts{
After nearly two decades of design, construction and commissioning, the CMS detector was operated with colliding LHC proton beams for the first time in November 2009. Collision data were recorded at centre-of-mass energies of 0.9 and 2.36 TeV, and analyzed with a fast turn-around time by the CMS collaboration. In this talk I will present a selection of commissioning results and striking first physics resonances observed. Then I will discuss the analysis of the transverse momentum and rapidity distribution of charged hadrons, which led to the first CMS physics publication. The excellent performance of the CMS detector and agreement with predictions from simulation are impressive for a collider detector at startup and show a great potential for discovery physics in the upcoming LHC run. 
}

\section{Introduction}

The CMS experiment~\cite{CMS} recorded the first LHC~\cite{LHC} proton-proton
collisions on Monday the 23$^{\rm rd}$ of November, 2009. In the weeks that 
followed, CMS collected approximately 350 thousand collision events 
at an energy of $\sqrt{s}$=0.9~TeV and 20 thousand events at 
$\sqrt{s}$=2.36~TeV with good detector conditions and the magnet 
switched on at the nominal value of 3.8~T. This corresponds 
to about 10~$\mu$b$^{-1}$ of integrated luminosity, close to 85\% of 
the collisions delivered by the LHC.

The recorded data sample is still many orders of magnitude too small
to do the physics studies for which CMS was designed. However, it is 
sufficient to assess the general quality and the proper 
functioning of the detector, the algorithms and the modeling of the 
detector response in the simulation. This is a crucial step in
the preparation of the experiment for physics. 

Section~\ref{sec:craft} briefly presents the status of CMS
at startup, followed by a summary of the first
physics performance results in Sections~\ref{sec:firstPeaks} 
to~\ref{sec:jetmet}. These performance results formed the basis 
for a timely publication of the first physics measurement with
collision data, one month before this conference~\cite{dndeta}, 
as discussed in Section~\ref{sec:dndeta}. This is followed by the conclusions in
Section~\ref{sec:conclusion}.

\section{CMS Status at Startup}
\label{sec:craft}

In the three years preceding the first LHC proton-proton collisions,
CMS recorded and analysed more than a billion events with muons from
various sources. Three cosmic ray runs in 2006, 2008 and 2009 
recorded about 300 million cosmic ray muon events each. Over a million
beam halo muons were recorded during LHC commissioning in 2008 and 2009, 
as well as more than a thousand so-called beam-splash events. These 
beam-splash events occur when LHC intentionally dumps a single bunch 
of the beam on a collimator about 150~m upstream from CMS, leading to 
a flood of muons traveling through the detector simultaneously.


Detailed analysis of these events resulted in crucial improvements in 
the alignment of the detector, modeling of the magnetic field, understanding
of the response of different subdetectors to muons, calibration,
noise characteristics and synchronization. The results of these 
studies are described in 23 performance papers~\cite{CRAFT}, submitted by
CMS just before the start of collisions in 2009. The papers have been 
accepted by JINST and are expected to appear in a single volume of the 
journal, soon after this conference.



The detector simulation that was thus tested and validated before 
collisions was used without further adjustments in the first CMS 
physics paper and for all other results presented in this
report. The only parameter that had to be adapted was the longitudinal
distribution of the primary collision vertices, which was tuned to match
the LHC operating conditions.

\section{The first Physics Resonance in Collisions}

\label{sec:firstPeaks}

\begin{figure}[!htbp]
\begin{center}
\includegraphics[width=\textwidth]{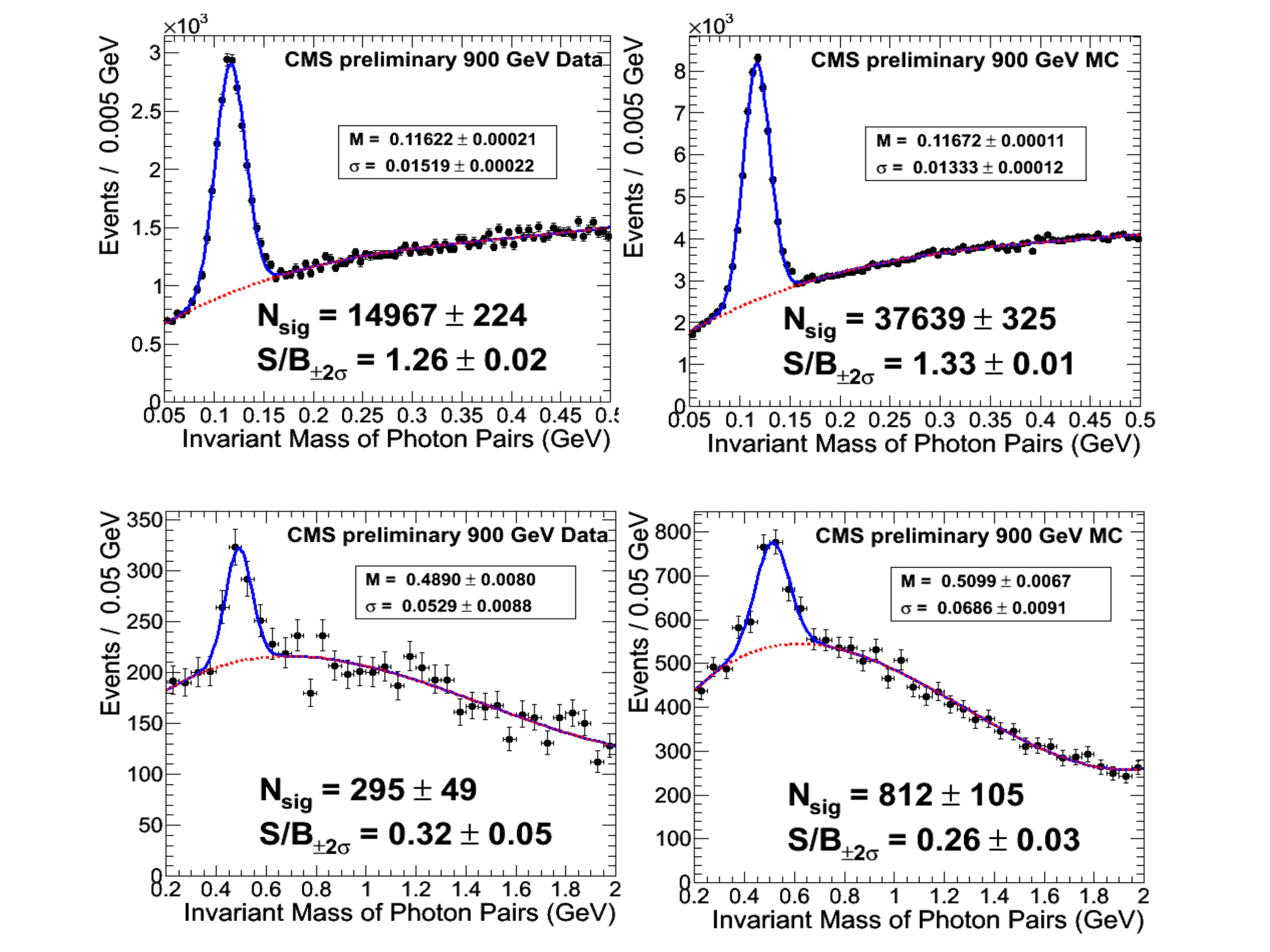}
\caption{Uncorrected photon-pair invariant mass peaks in data (left) and simulation (right) in the region
         around the mass of the $\pi^0$ (top) and $\eta$ (bottom) resonances.}
\label{fig:firstPeaks}
\end{center}
\end{figure} 

\begin{figure}[!htbp]
\begin{center}
\includegraphics[width=0.35\textwidth,angle=90]{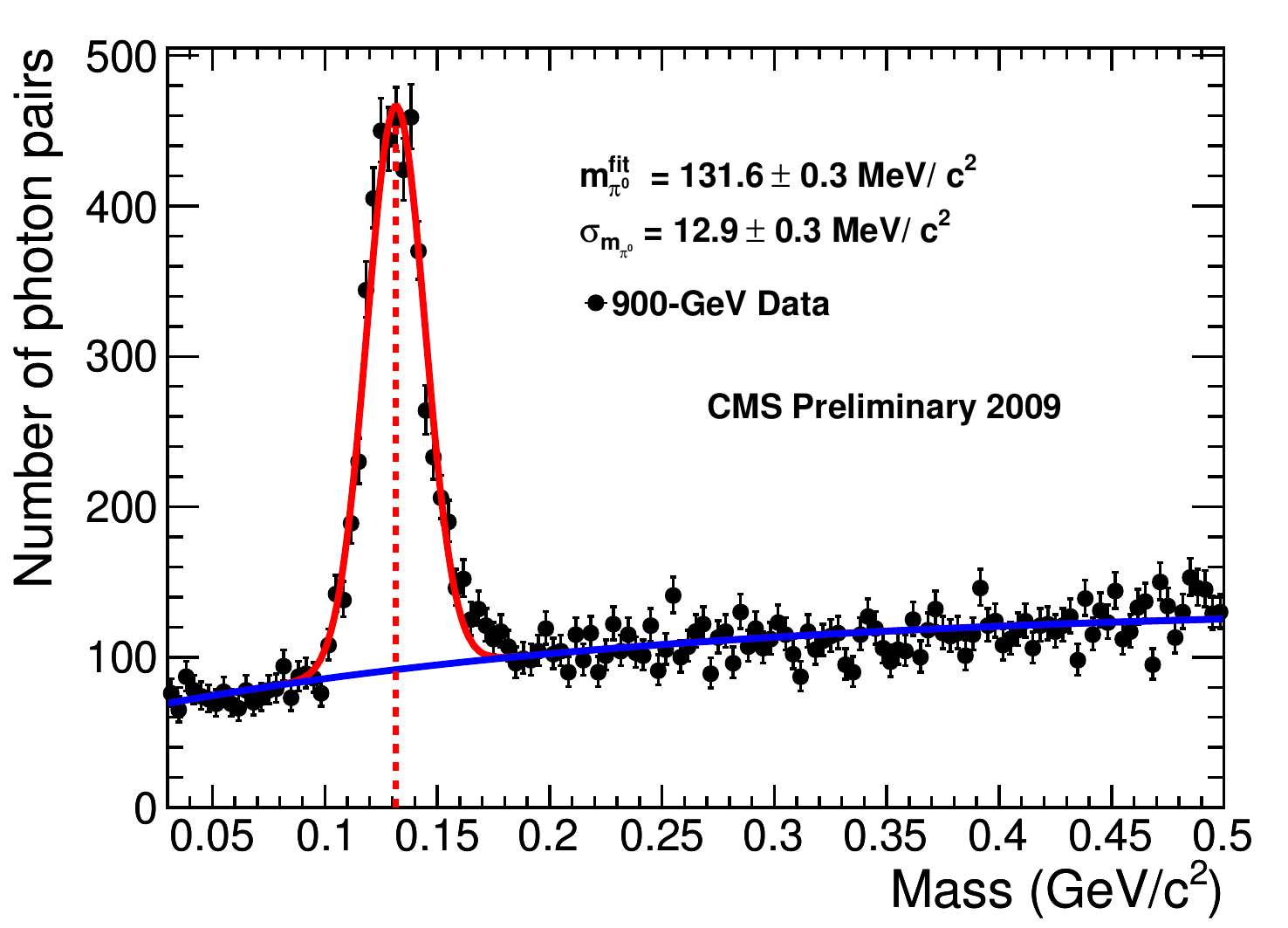}
\includegraphics[width=0.45\textwidth,angle=90]{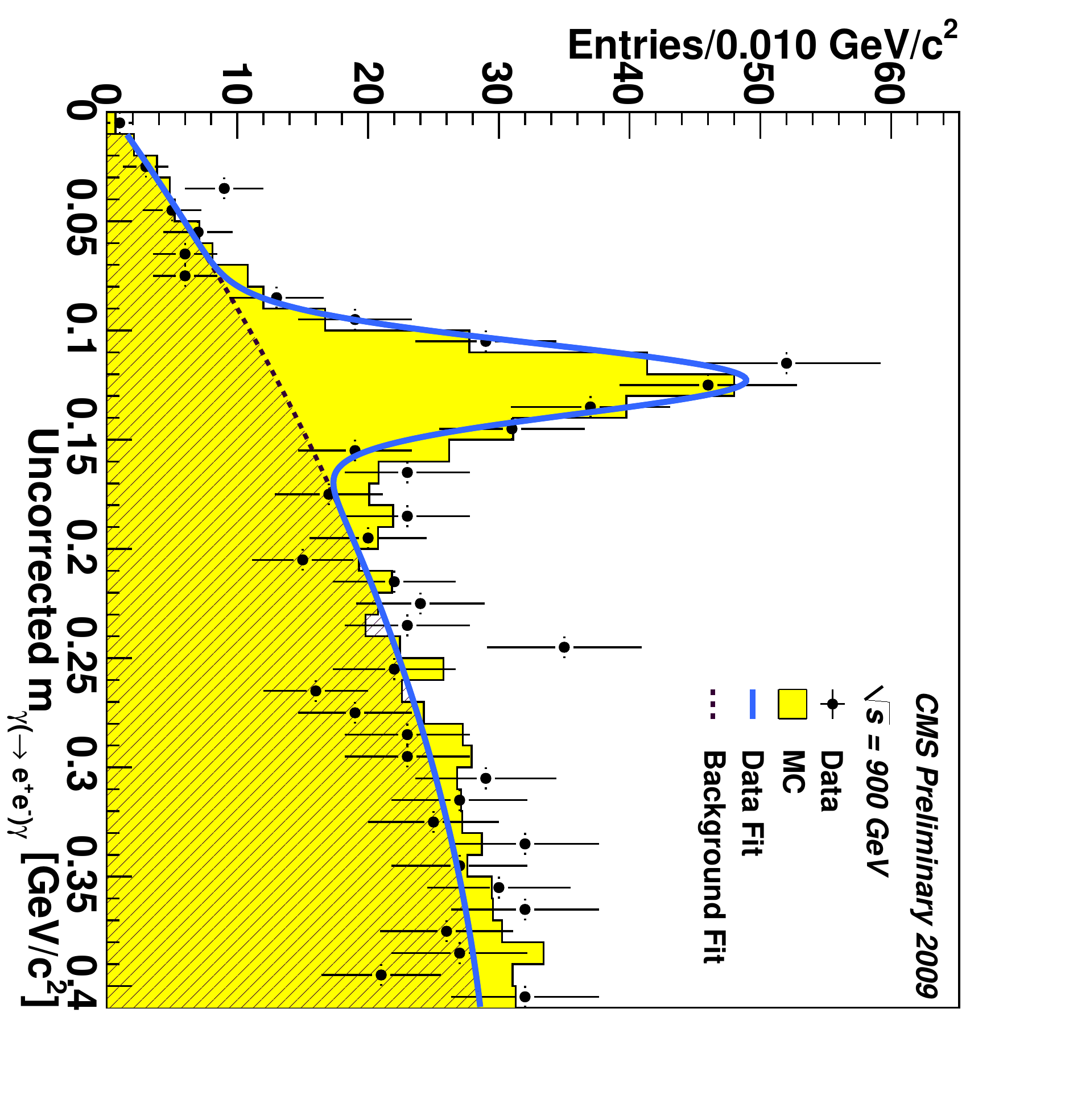}
\caption{Left: Photon-pair invariant-mass distribution in the barrel ($| \eta | <$ 1.0) for the 
  data. The $\pi^{0}$ mass peak is fitted with a Gaussian (red line) and
  the combinatorial background is described with an exponential function (blue line). 
  Right: Invariant mass of $\pi^0$ candidates reconstructed as one photon and an e$^+$ e$^-$ pair.
  Data (black points) are superimposed on the MC expectation (filled yellow histogram), 
  normalized to the number of selected conversion candidates. A 
  fit to the data points, obtained with a Gaussian function summed with a second 
  order polynomial to describe the combinatorial background component, is also shown.}
\label{fig:pi0challenge}
\end{center}
\end{figure}

The first physics resonance observed in collision data was the di-photon invariant mass from
the decay of neutral pions $\pi^0 \ra \gamma \gamma$, visible even 
after the first run with 191 recorded events when the experimental magnets were
still switched off. The peak, together with a dedicated simulated sample
(with magnetic field off) was ready and approved for public presentation within
three days after the first collisions. Updated versions of the $\pi^0$ 
di-photon invariant mass plots~\cite{DPS-10-001}, with more data and simulated events, 
are shown in Fig.~\ref{fig:firstPeaks}(top). For this plot only photon candidates
in the barrel ($|\eta|<$1.479) are used, requiring basic shower shape cuts, a 
transverse photon energy $E_{\rm T}$ above 300~MeV, and the transverse momentum $p_{\rm T}$ of the 
reconstructed $\pi^0$ above 900~MeV.
Similarly, the eta resonance $\eta \ra \gamma \gamma$ is shown for
data and simulation in Fig.~\ref{fig:firstPeaks}(bottom). In this case a photon $E_{\rm T} >$ 400~MeV 
and $\eta$ $p_{\rm T} >$ 2 GeV was required.  
The masses shown are based on the measured
photon energies without corrections for shower containment and 
energy lost before the calorimeter, which explains why the observed masses are a few percent below the
known $\pi^0$ and $\eta$ mass. However, the mass reconstructed in data and simulation
agrees to within about 2\% even at this relatively low energy, 
in agreement with the expected calibration at start-up. When applying a simulation-based 
correction for single-photon energies~\cite{PFT-10-001}, the mass moves to within 2\% of
the PDG~\cite{PDG} value, as shown in Fig.~\ref{fig:pi0challenge}(left).
The $\pi^0$ mass peak is also shown for events where one of the photons converted in the tracker
and is reconstructed as an e$^+$ e$^-$ pair.~\cite{EGM-10-001}

\section{Tracking}
\label{sec:tracking}

The CMS silicon tracker and tracking algorithms have performed excellently from the
start of data taking. Beam spot and primary vertices are reconstructed
with high efficiency and resolution close to the expectation from 
simulation~\cite{TRK-10-001}. Within hours after the first run with magnetic field
switched on, invariant mass peaks were reconstructed of the decays of the
neutral kaon $K_S^0 \to \pi^+\pi^-$ and $\Lambda^0 \to p\pi^-$ (and their 
charge conjugates), with a 
mass scale correct to better than 0.1\%, and 
good agreement between data and simulation in resolution. The agreement
in mass at low $p_{\rm T}$ is a new, independent, confirmation that the 
scale of the magnetic field is modeled accurately, and also provides
a first indication that the  
description of material effects in the tracker is realistic.

As these particles are long-lived ($c\tau > 1$~cm) and decay to a pair
of charged particles, they provide a so-called $V^0$ signature, consisting
of two oppositely charged tracks which are detached from the primary vertex 
and form a good vertex. To ensure good track quality, a track is required
to have more than 5 hits, a normalized
$\chi^2$ less than 5, and a transverse impact parameter with respect to the beamspot
greater than $0.5 \, \sigma_{\rm IP}$ where $\sigma_{\rm IP}$ is the calculated uncertainty (including
beamspot and track uncertainties).  The reconstructed $V^0$ decay vertex must have a $\chi^2$ less than 7
and a transverse separation from the beamspot greater than 15~$\sigma_{\rm T}$ where $\sigma_{\rm T}$
is the calculated uncertainty. If either of the daughter tracks have hits that
are more than 4$\sigma_{3D}$ inside the $V^0$ vertex (toward the primary vertex), 
the $V^0$ candidate is discarded.

\begin{figure}[hbtp]
  \begin{center}
    \includegraphics[angle=90,width=0.49\textwidth]{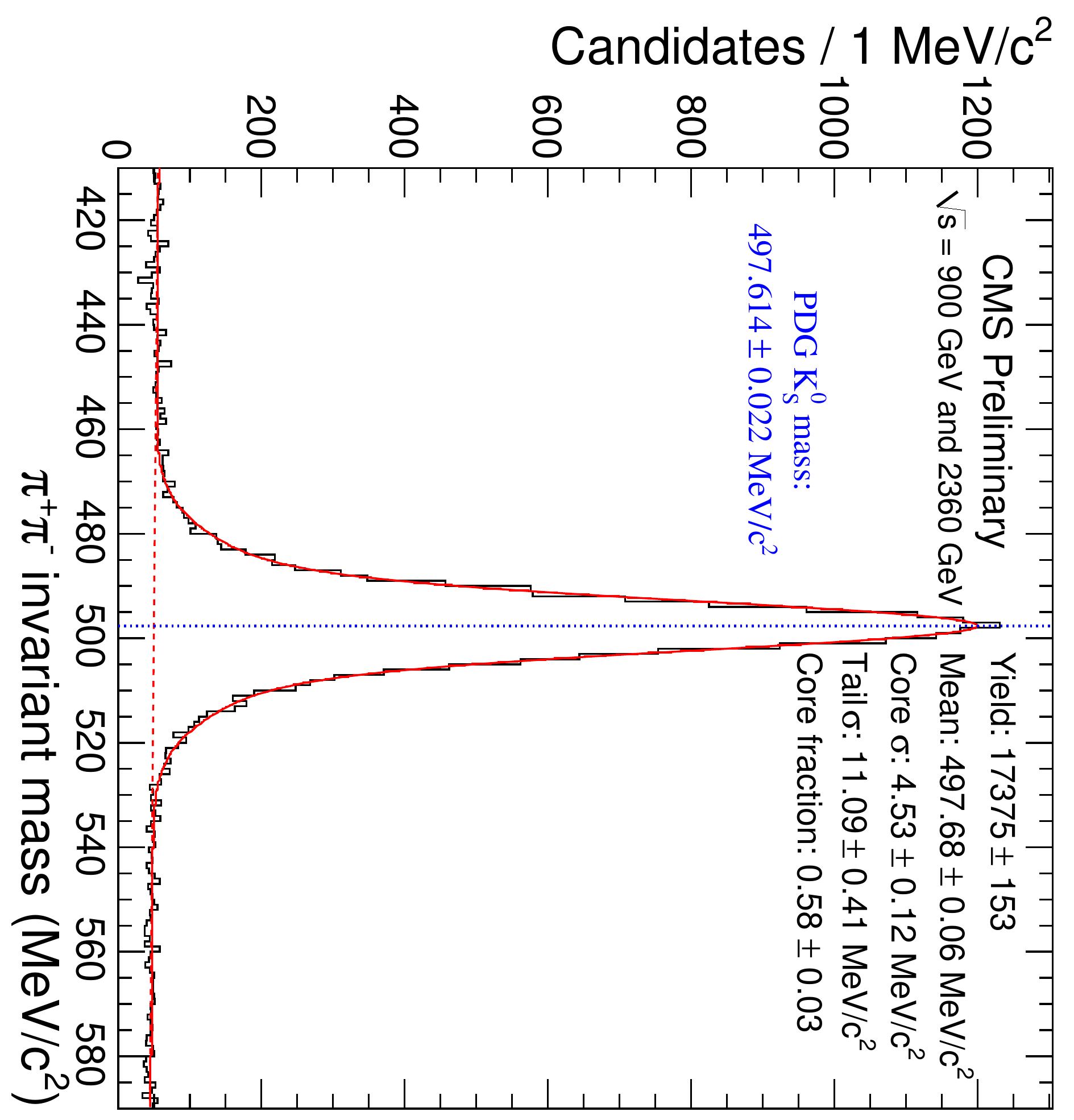}
    \includegraphics[angle=90,width=0.49\textwidth]{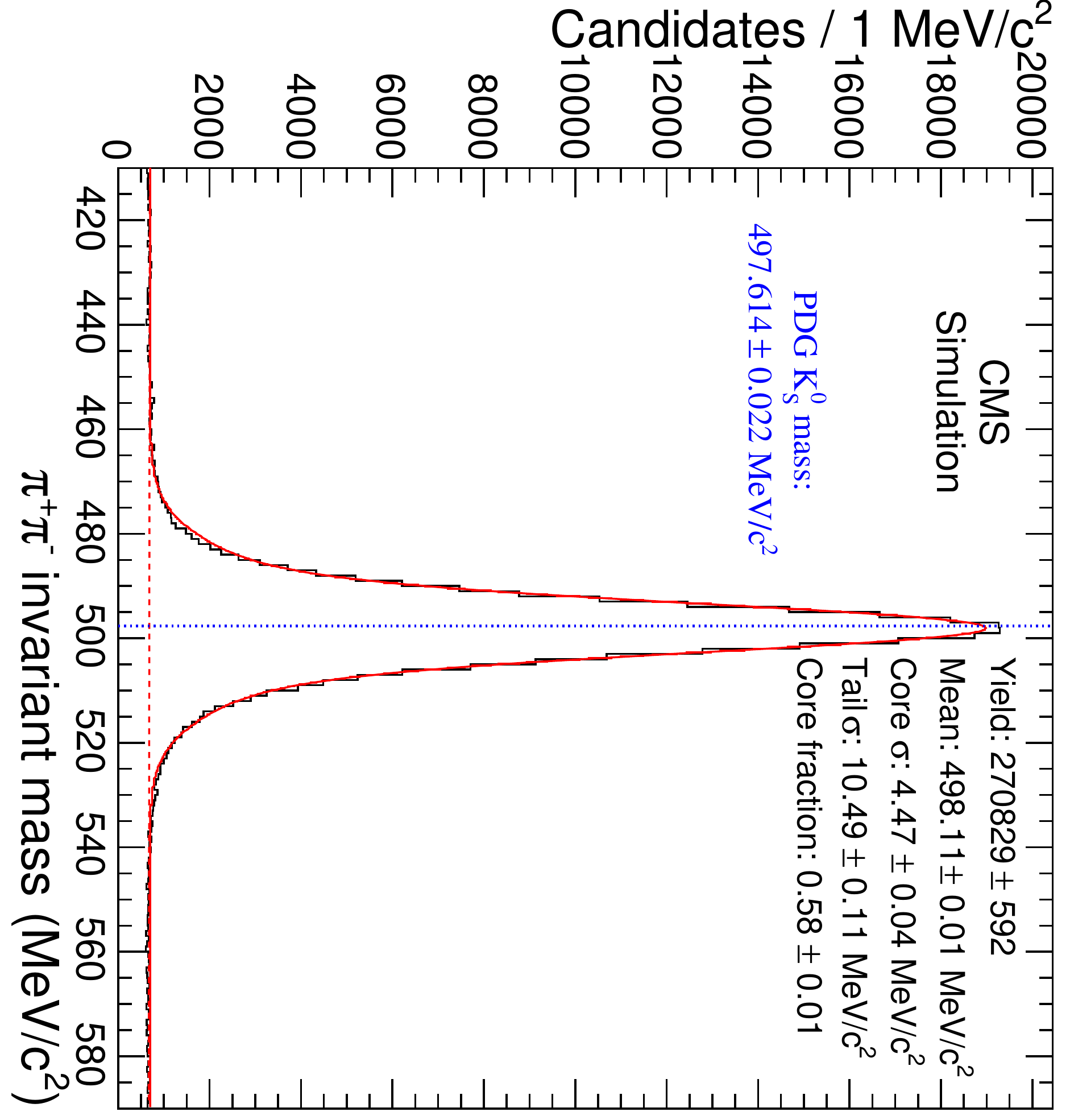}
    \includegraphics[angle=90,width=0.49\textwidth]{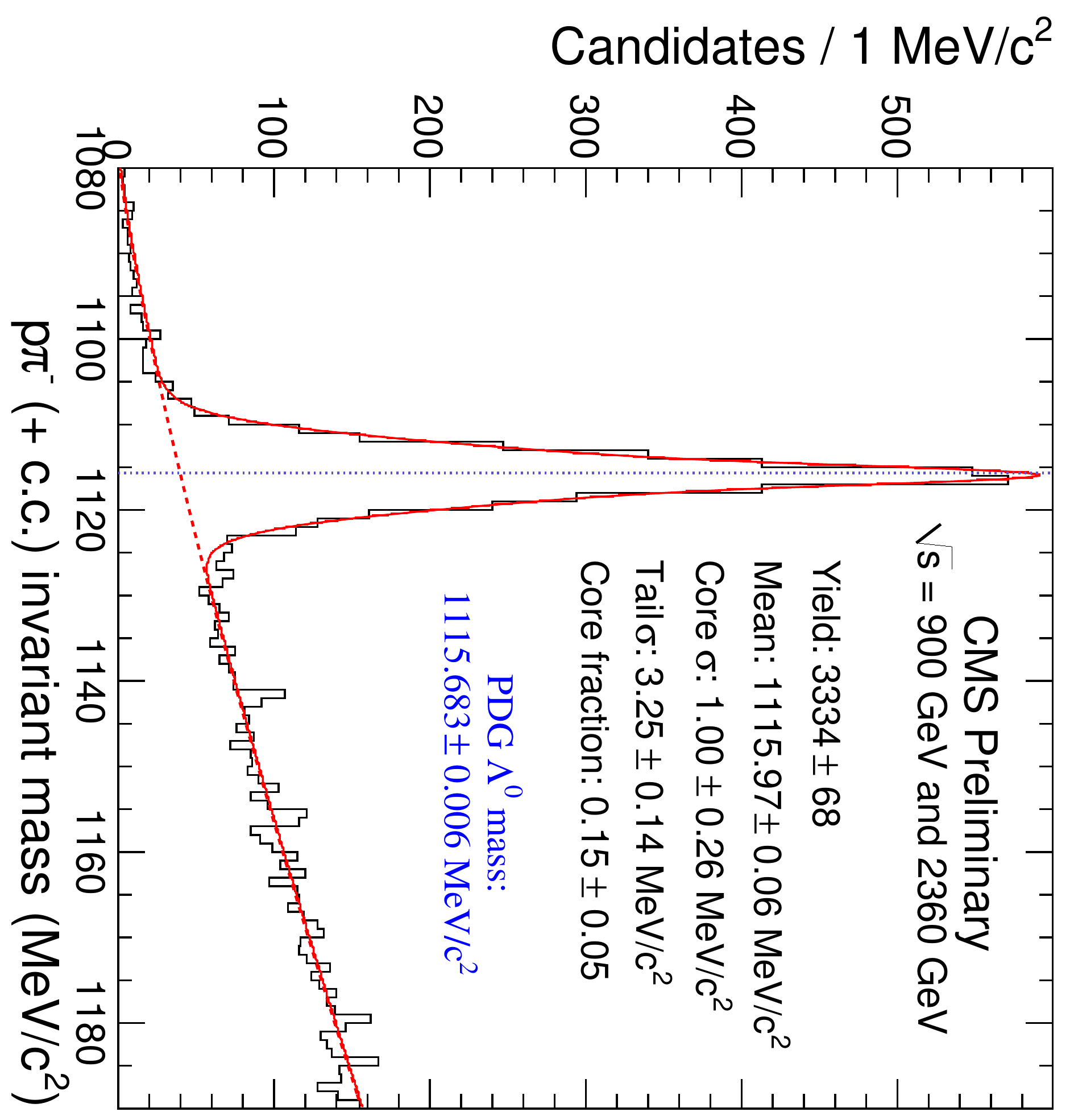}
    \includegraphics[angle=90,width=0.49\textwidth]{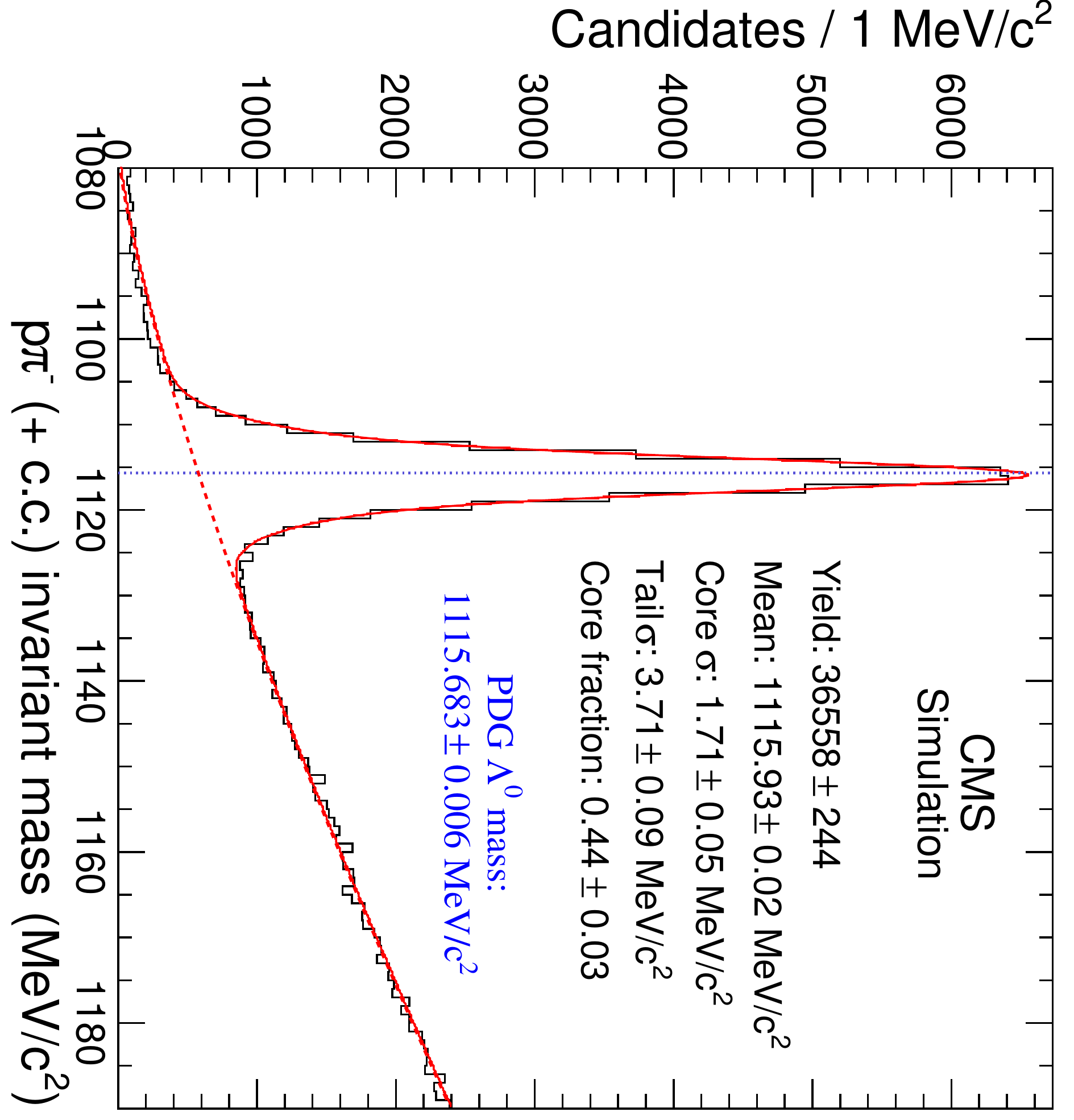}
    \caption{Fitted $\pi^+\pi^-$ mass for data (top left) and simulation (top right). Fitted $p\pi^-$ (+ charge conjugate) mass for data (bottom left) and simulation (bottom right).  
Uncertainties shown are statistical only.}
    \label{fig:masses}
  \end{center}
\end{figure}

The mass resolution of the $V^0$ depends on $\eta$ as well as the decay vertex position and 
a single Gaussian was not a sufficiently accurate functional form for the signal.  Therefore,
a double Gaussian with the same mean was used to fit the signal.  For the background shapes,
a linear background was used for $\pi^+\pi^-$ and the function $a(m-m_p-m_\pi)^b$
was used for the $p\pi^-$ spectrum.  
The $\pi^+\pi^-$ and $p\pi^-$ mass distributions, along with the overlaid
fits, are shown in Fig.~\ref{fig:masses}.

\subsubsection{Reconstruction of the $K^*(892)^-$, $\Xi^-$ and $\phi$ Resonances}

The reconstructed sample of $V^0$ particles is exploited to search for other particles as well.

\begin{figure}[hbtp]
\begin{center}
	  \includegraphics[width=0.54\linewidth]{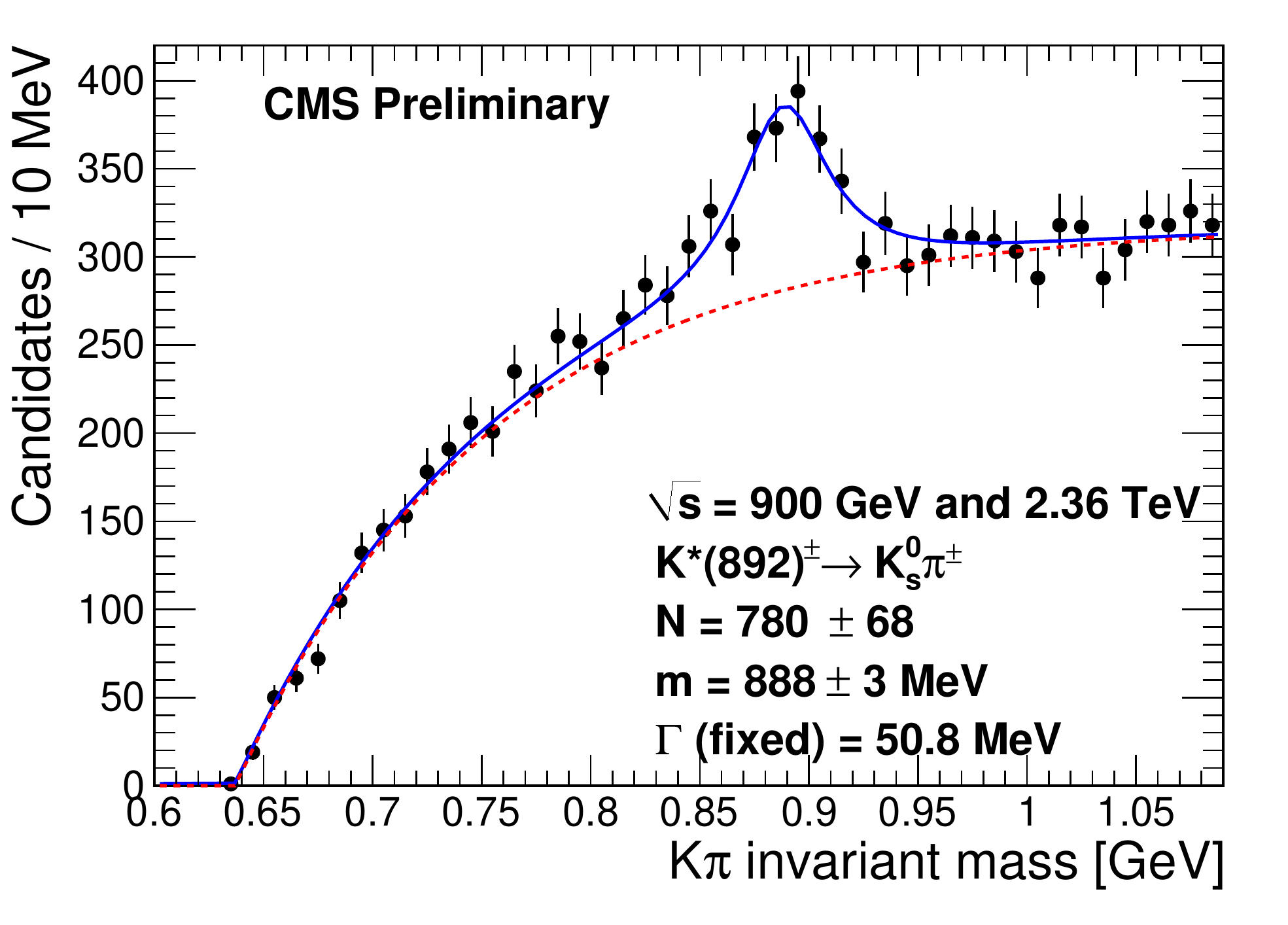}
	  \includegraphics[angle=90,width=0.4\linewidth]{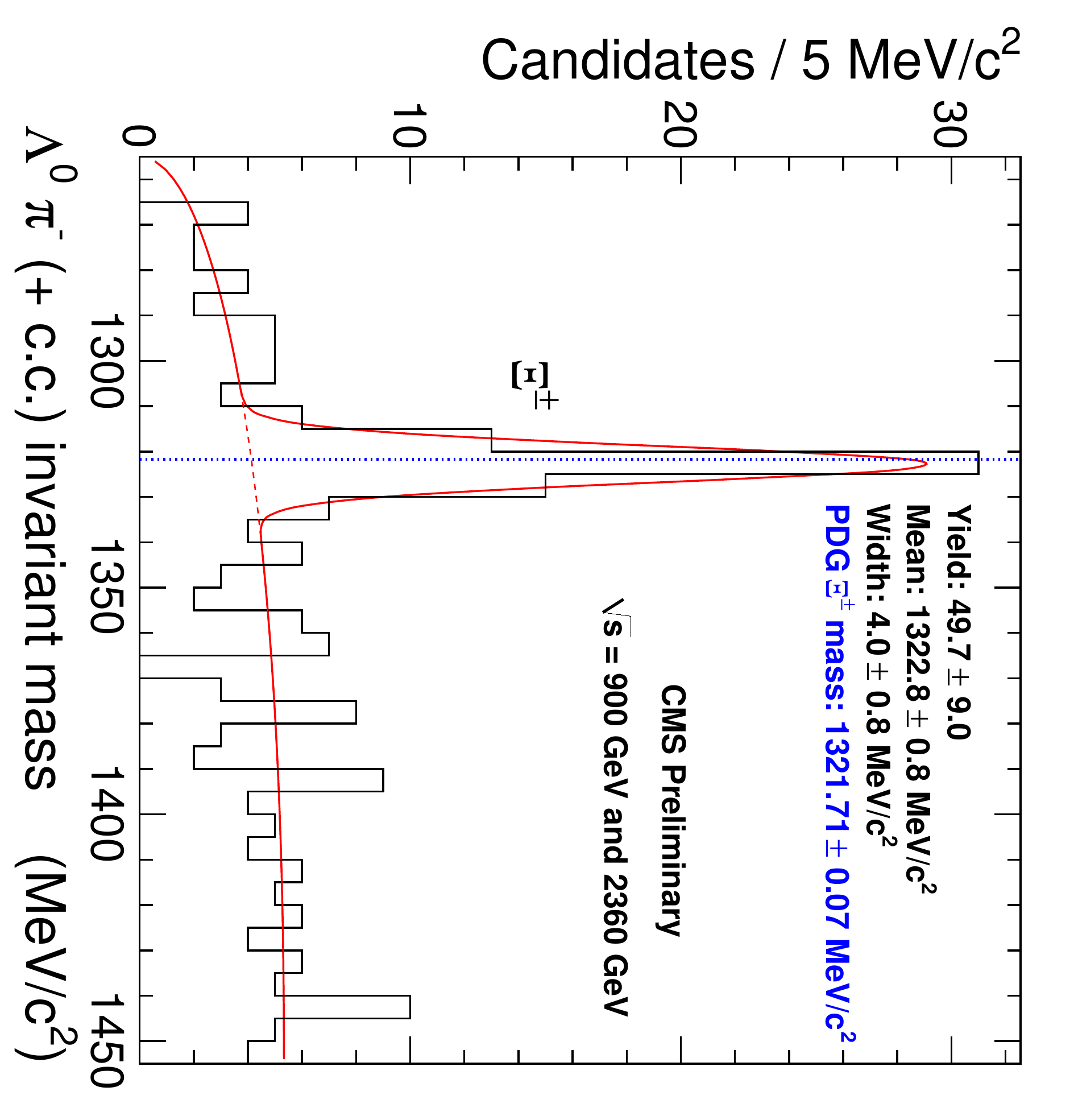}
\caption{Left: $K_S^0 \pi^-$ invariant mass plot with a fit for the $K^*(892)^-$. Right:
$\Lambda^0\pi^-$ invariant mass plot with a fit for the $\Xi^-$.}
\label{fig:XiKstar}
\end{center}
\end{figure}

First, $K_S^0$ candidates are combined with charged tracks from the primary vertex to
search for the strong decay $K^*(892)^- \to K_S^0\pi^-$.  
The $K_S^0$ candidates must have a mass within 
20~MeV of the nominal PDG mass and the $K_S^0$ flight path must pass within 2~mm of the primary vertex.  
The $K_S^0\pi^-$ invariant mass is calculated using the PDG value of the
$K_S^0$ mass and is shown in Fig.~\ref{fig:XiKstar}(left).  The figure also shows an overlay
of a fit to the $K_S^0\pi^-$ mass distribution.  The fit uses a relativistic Breit-Wigner
for the signal plus a threshold function for the background. More details are given elsewhere~\cite{TRK-10-001}.
The mass returned by the fit is $888 \pm 3$~MeV, consistent with the world average 
value of $891.66 \pm 0.26$~MeV~\cite{PDG}.

The second particle, the $\Xi^-$, was reconstructed through its decay to
$\Lambda^0\pi^-$.  As the $\Xi^-$ is a long-lived baryon, the topology of this decay is
different than the $K^*(892)^-$.  The $\pi^-$ from the $\Xi^-$ decay will be detached
from the primary vertex rather than originating from the primary vertex. 
$\Lambda^0$ candidates
with a mass within 8~MeV of the PDG value were combined with charged tracks with the
same sign as the pion in the $\Lambda^0$ decay (the track with the smallest $p_{\rm T}$).  
The resulting mass plot is shown in Fig.~\ref{fig:XiKstar}(right). 
The measured mass of $1322.8 \pm 0.8$~MeV is in agreement with the world
average value of $1321.71 \pm 0.07$~MeV~\cite{PDG}.

Finally, a search for the $\phi$(1020)$ \to K^+K^-$  resonance was performed, exploiting
the possibility to separate kaons and pions at
low $p_{\rm T}$ by measuring $dE/dx$ in the silicon detector layers. This 
analysis was described in a dedicated contribution to this conference~\cite{TRK-10-001,PhiKK}.
%
%
Also in this case the mass that is fitted, $1019.58 \pm 0.22$~MeV, is in good agreement with
the PDG value of the $\phi$ mass of $1019.455 \pm 0.020$~MeV.~\cite{PDG}  
%
%
%
To conclude, all
five resonances observed were found to have masses in agreement (to better
than 0.1\%) with the values listed in the PDG.

\begin{table}[hbt]
\caption{Absolute and relative mass bias of several mass resonances observed in early CMS data, using track reconstruction and simulation out-of-the-box without any corrections. Only statistical uncertainties are included for the CMS results.\label{tab:mass}}
\vspace{0.4cm}
\begin{center}
\begin{tabular}{|l|c|c|c|c|c|}
\hline
          & $\ko_S$ &  $\Lambda^0$ & $\Xi^{\pm}$ & $K^{* \pm}$ & $\phi$ \\
\hline
Mass Bias (GeV) & & & & & \\
$\Delta m = $ & & & & & \\
$(m_{\rm data}-m_{\rm PDG})-$ & -0.37$\, \pm \,$0.07 &  0.04$\, \pm \,$0.06 & 0.0$\, \pm \,$0.9 & -4.0$\, \pm \,$3.1 & -0.22$\, \pm \,$0.26 \\
$(m_{\rm MC}-m_{\rm gen})$ & & & & & \\
\hline
Relative bias (\%) & &  & & & \\
$\Delta m / m_{\rm PDG}$ & -0.074$\, \pm \,$0.014 &  0.004$\, \pm \,$0.005 & 0.00$\, \pm \,$0.07 & -0.5$\, \pm \,$0.4 & -0.02$\, \pm \,$0.03 \\
\hline
\end{tabular}
\end{center}
\end{table}

\subsubsection{Basic b-tagging Observables}
\label{sec:btag}

\begin{figure}[hbtp]
  \begin{center}
    \includegraphics[width=0.49\linewidth]{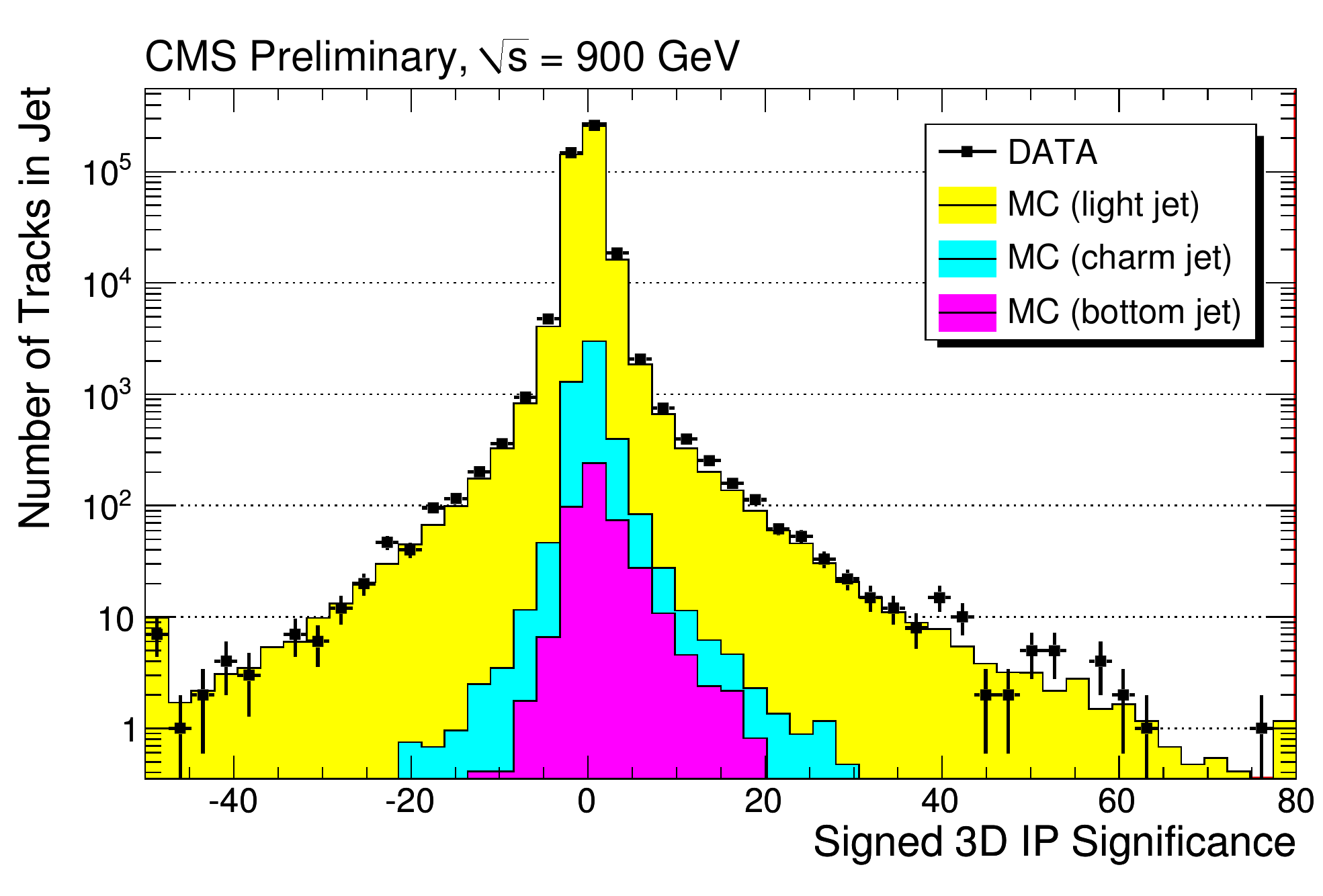}
    \includegraphics[width=0.49\linewidth]{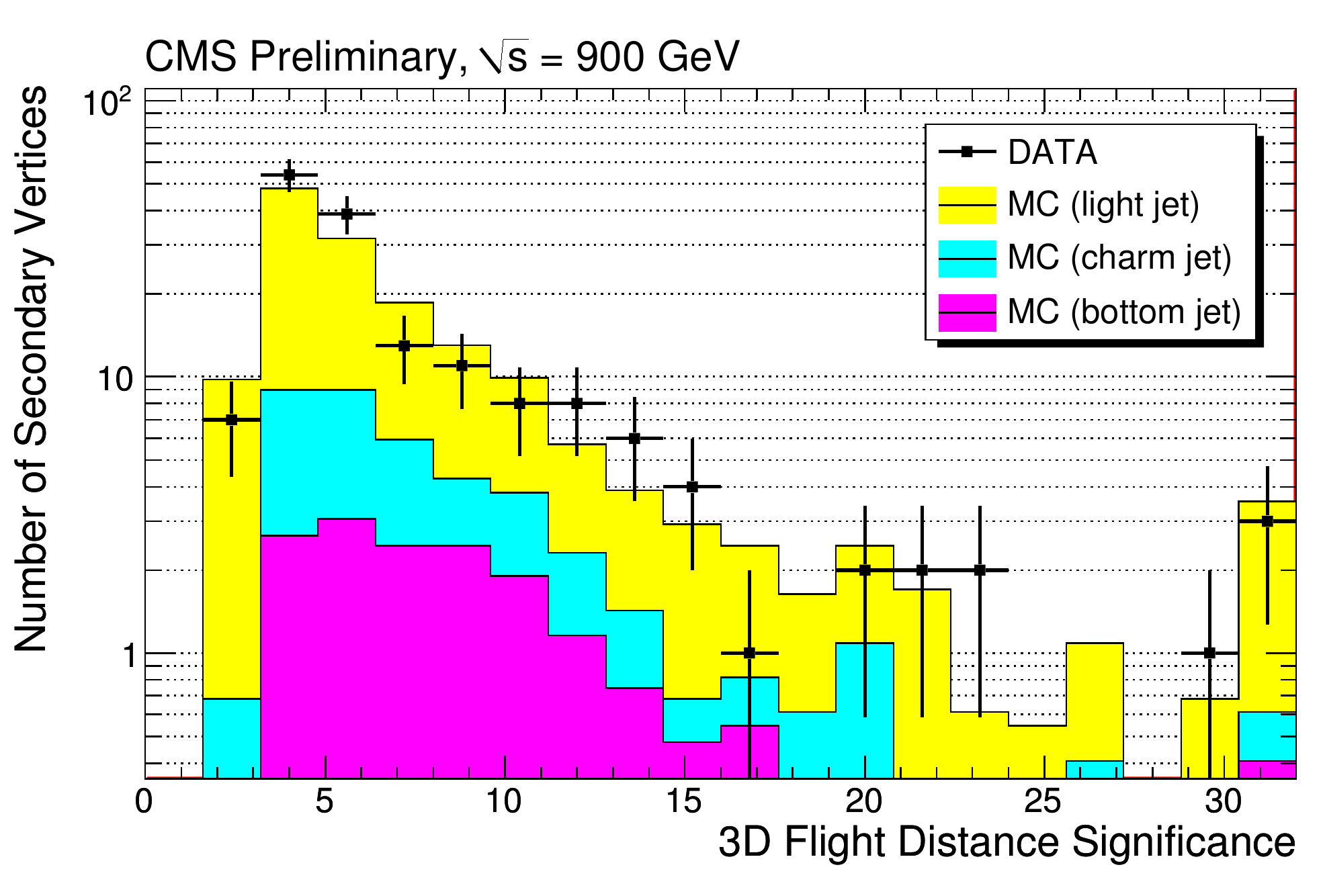}
    \includegraphics[width=0.49\linewidth]{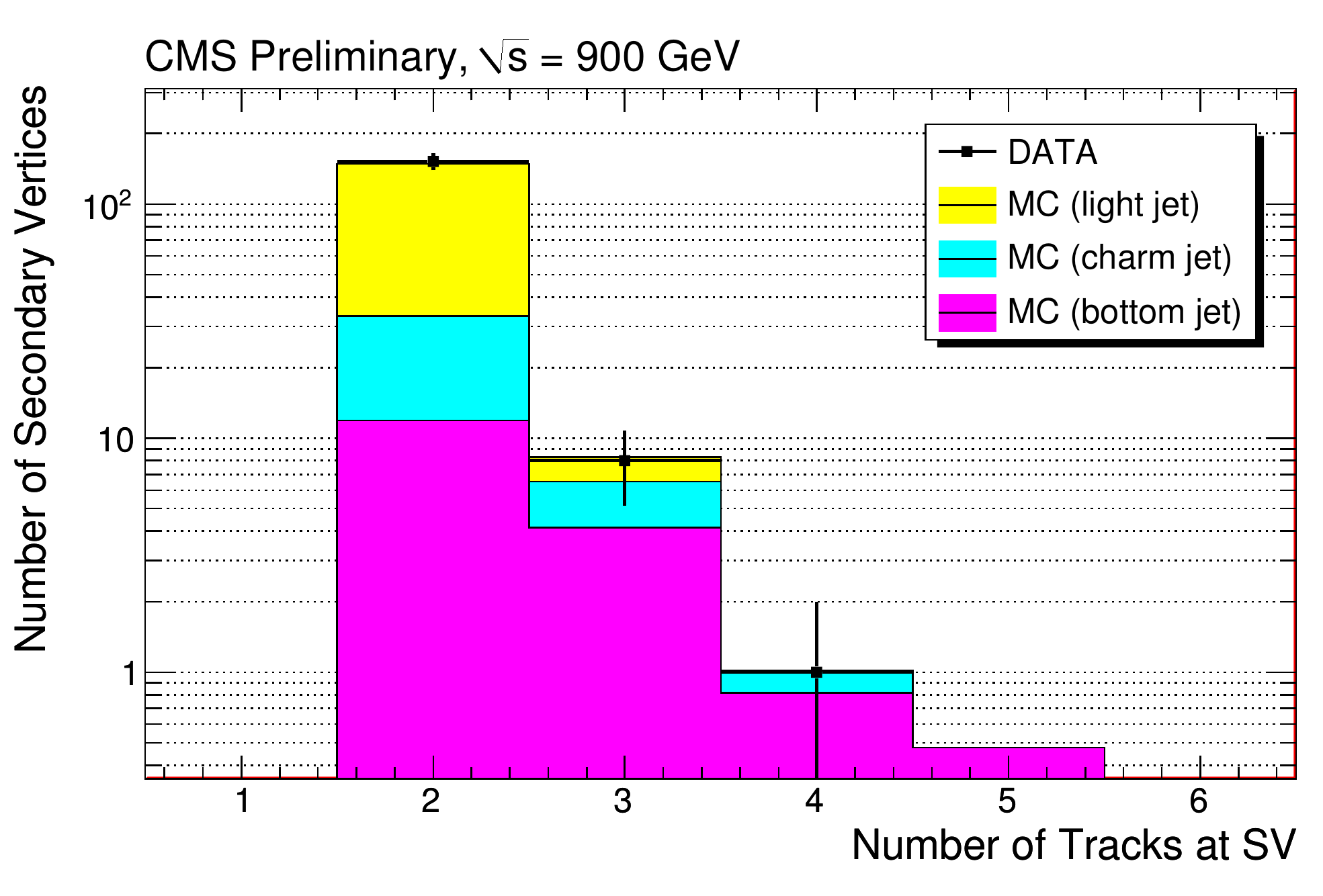}
    \hglue 2.4cm \includegraphics[width=0.33\linewidth]{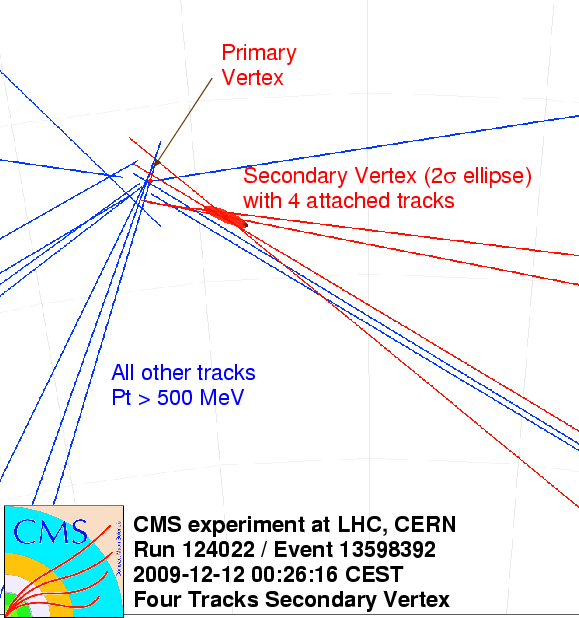}
    \caption{Distribution of the significance of the three-dimensional impact parameter for all tracks 
      in the jet (top left) and the significance of the three-dimensional displacement of the secondary vertex (top right).  
      The data is shown as black dots while for the simulation the contributions from light flavor, 
      charm, and bottom are shown as filled histograms. The two outermost bins contain the respective histogram overflow. 
      The number of tracks in a secondary vertex (bottom left), and the only event with a 4-track secondary vertex (bottom right).}
    \label{fig:btag}
  \end{center}
\end{figure}

A proper understanding of impact-parameter resolution of tracks and the 
reconstruction of secondary vertices is important for b-tagging, and
presents the next challenge for the tracker and tracking algorithms.
To validate the basic ingredients for b-tagging on a larger sample of events, 
a few changes to the reconstruction chain were applied~\cite{TRK-10-001} with respect to the
default algorithm~\cite{BTV-09-001}, relaxing the requirements on the 
acceptance of tracks, jets, and the matching between them.  
Figure~\ref{fig:btag}(top left) shows the three-dimensional
impact parameter significance distribution for all tracks in a selected sample of 
jets, computed with respect to the reconstructed primary vertex.

The secondary vertex reconstruction using the tracks associated to jets has also been 
slightly modified~\cite{TRK-10-001}, with relaxed requirements. To 
suppress $K_S^0$ candidates, the transverse secondary vertex separation was required to be
less than 2.5~cm and the secondary vertex invariant mass more than 15~MeV from the nominal $K_S^0$ mass.
The vertex flight distance is compared to what is expected from
a simulation of minimum bias events in Fig.~\ref{fig:btag}(top right). 
In general the agreement between data and simulation over the entire range of interest for the 
variables considered is remarkable. While many two- and three-track vertices are
reconstructed, only one four-track vertex is expected (see Fig.~\ref{fig:btag}, bottom left) and one is 
found in the data~\cite{DPS-10-003}. The corresponding event display is shown in the same figure (bottom right).  

\section{Unification of Calorimetry and Tracking: the Particle Flow Approach}
\label{sec:PFlow}

With its combination of a strong magnetic field, precise silicon tracker and an electromagnetic
calorimeter with fine lateral segmentation the CMS design lends itself beautifully
for the Particle Flow approach. In this approach one aims at reconstructing all 
stable particles in the event (i.e., electrons, muons, photons and charged and neutral hadrons) from the 
combined information from all CMS sub-detectors,
to optimize the determination of particle types, directions and energies. Simulation studies have
shown~\cite{PFT-09-001} 
that in the case of CMS this can lead to an improvement of about a factor two 
in resolution for jets at low $p_{\rm T}$ ($<$50 GeV) and for missing transverse energy. 

A key ingredient is the linking of tracks with corresponding energy 
clusters in the calorimeters. This was the first aspect to focus
on once collision data became available. The angular matching between tracks
and calorimeter deposits was shown to be reproduced very well by simulation~\cite{PFT-10-001}.

\begin{figure}[hbtp]
  \begin{center}
    \includegraphics[width=0.8\textwidth]{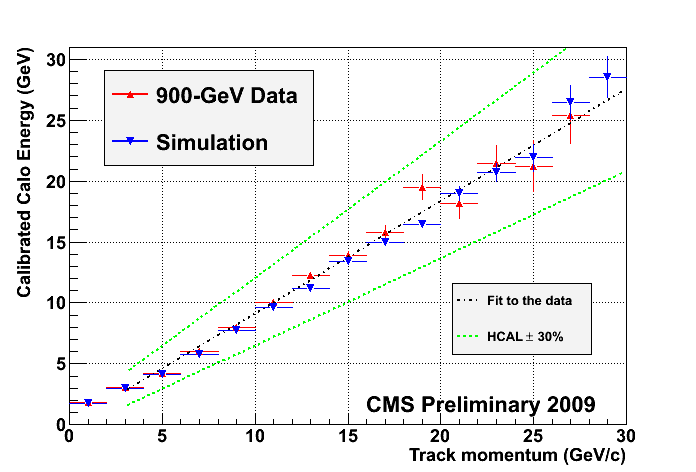}
    \caption{Average calorimeter response as a function of the track momentum for the 
900 GeV data (red upwards triangles with error bars) and the corresponding minimum-bias simulation 
(blue downwards triangles with error bars), integrated over the full tracker acceptance ($\vert \eta \vert < 2.4$).  
The dash-dotted line is a linear fit to the data, and the dashed lines show the same fit with a HCAL raw response
changed by $\pm 30\%$, to guide the eye.}
  \label{fig:PFlowHCAL}
\end{center}
\end{figure}

Once tracks and calorimeter clusters are matched in angle (or position), the measured track momentum can
be compared to associated calorimeter energy for the combined electromagnetic (ECAL)  
and hadronic (HCAL) calorimeter clusters, as shown as a function of the track $p_{\rm T}$ in Fig.~\ref{fig:PFlowHCAL}. 
Again data and MC simulation agree well. Since the ECAL energy scale was shown to be correct within 2\% and the tracking
momentum scale within 0.1\%, one can derive from this plot that the HCAL response simulation is correct to better than 5\%.

\section{Jets and Missing Transverse Energy}
\label{sec:jetmet}

In previous sections we have checked the calibration of all elements 
of tracking and calorimetry. These are used as ingredients for jet
reconstruction. CMS uses the anti-$k_{\rm T}$ clustering
algorithm~\cite{JME-10-001} with a cone size of 0.5 for commissioning, 
with three different types of inputs: {\em Calo Jets} are purely 
based on energy deposits in the calorimeter; {\em Track Jets} start 
from calo jets and use track information to improve the jet resolution;
and {\em Particle-Flow Jets} use all particles reconstructed by the 
Particle Flow algorithm as input.

\begin{figure}[hbtp]
  \begin{center}
    \includegraphics[width=0.32\textwidth]{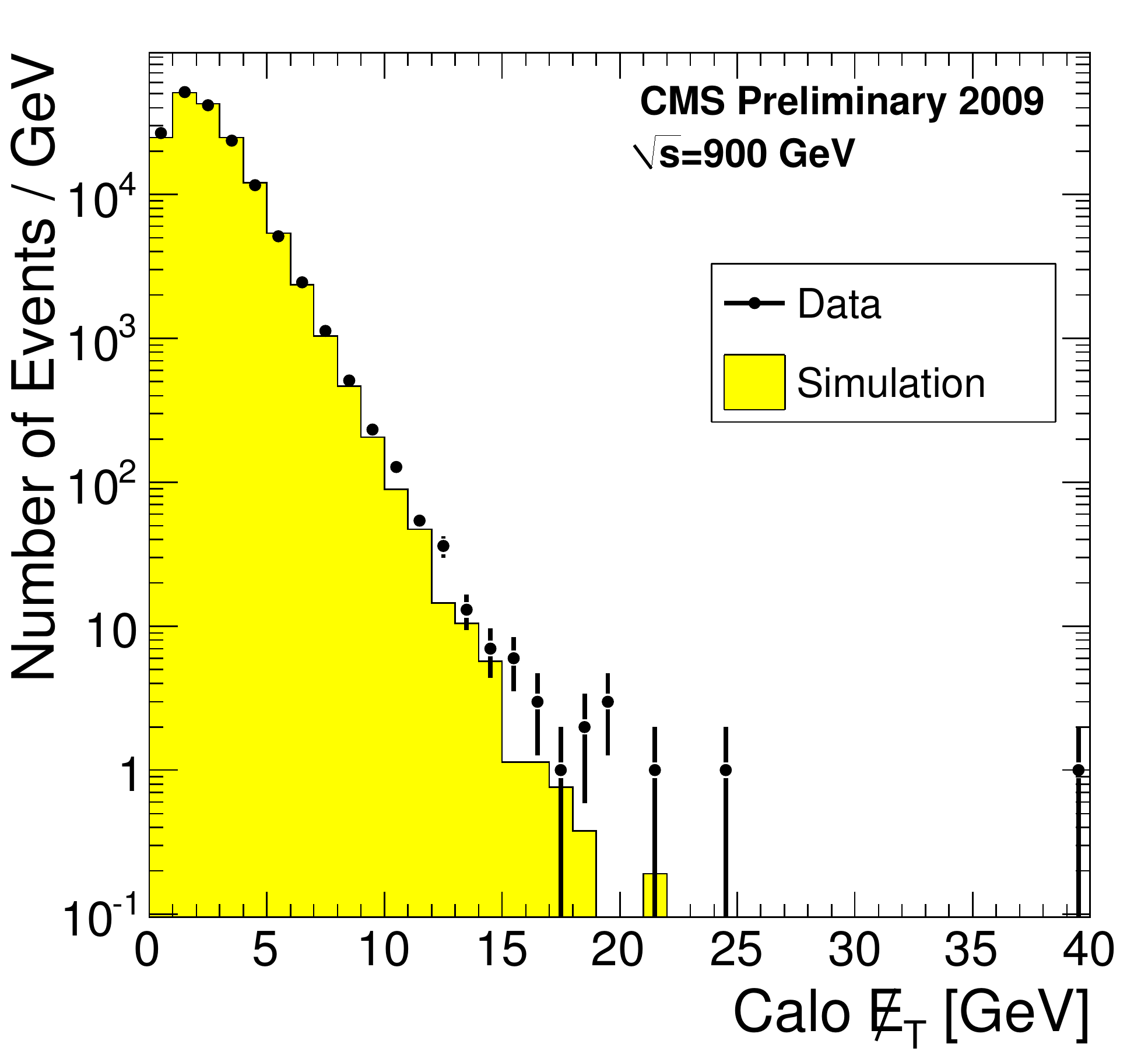}
    \includegraphics[width=0.32\textwidth]{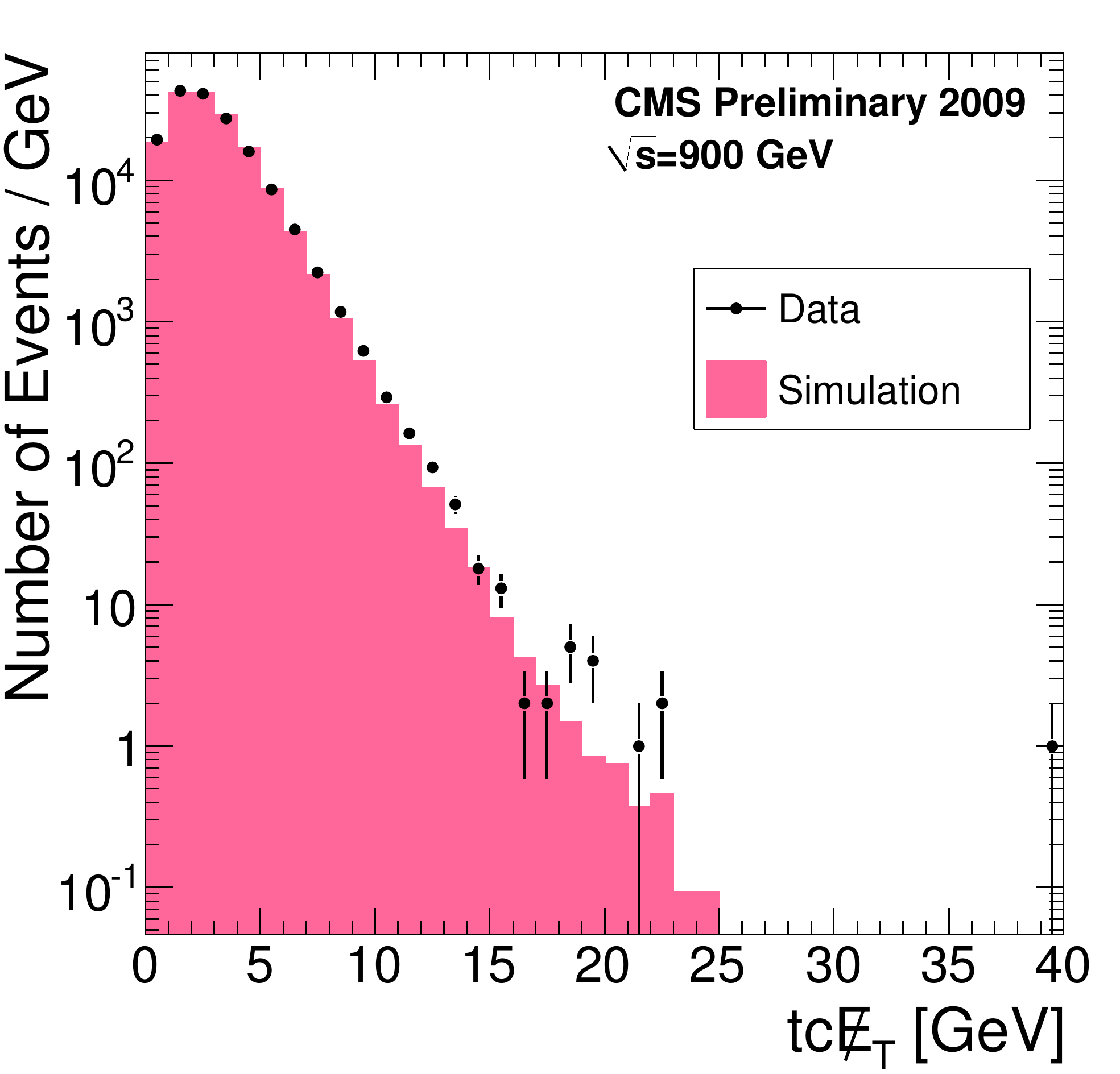}
    \includegraphics[angle=90,width=0.32\textwidth]{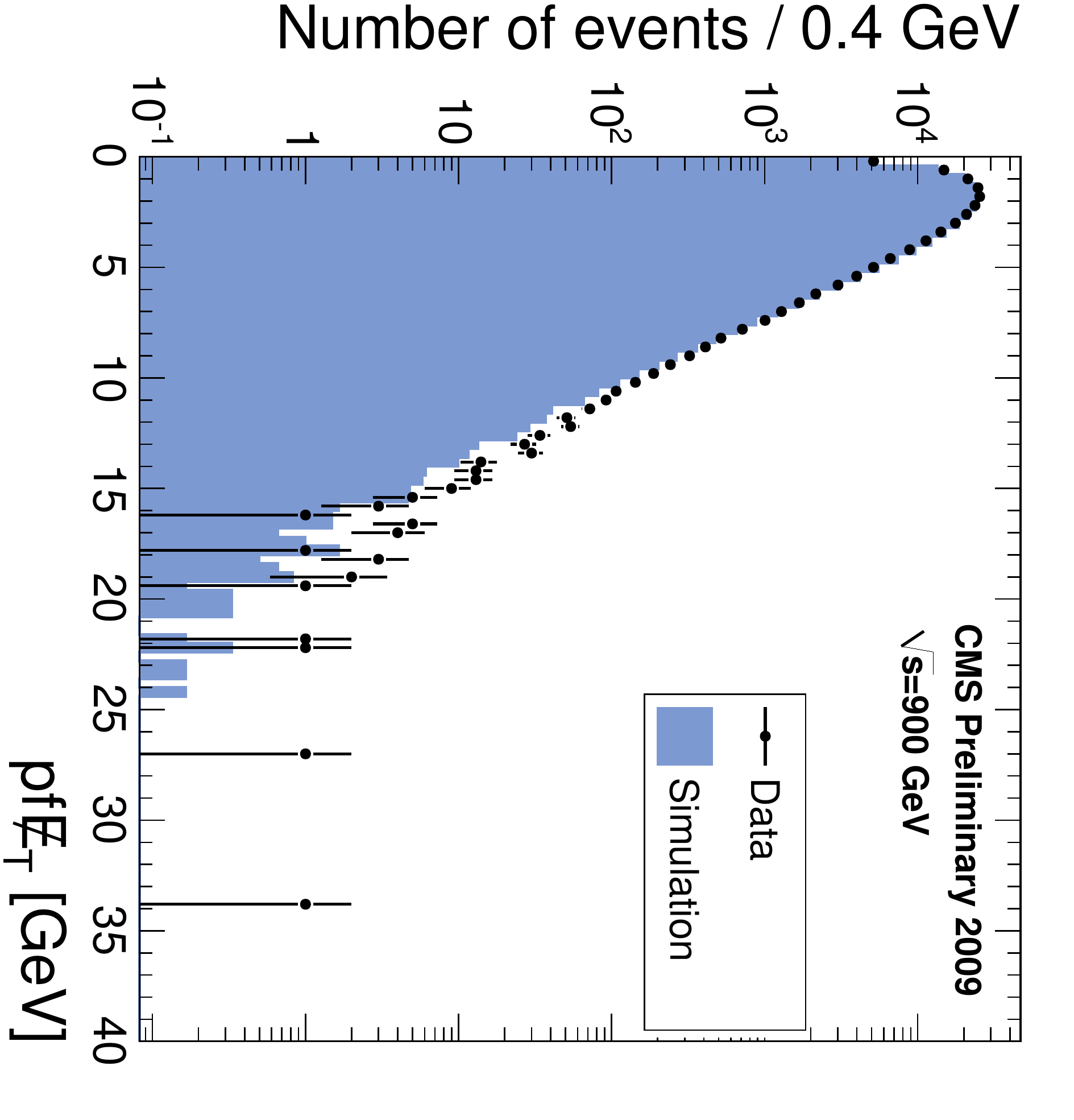}
    \includegraphics[angle=90,width=0.49\textwidth]{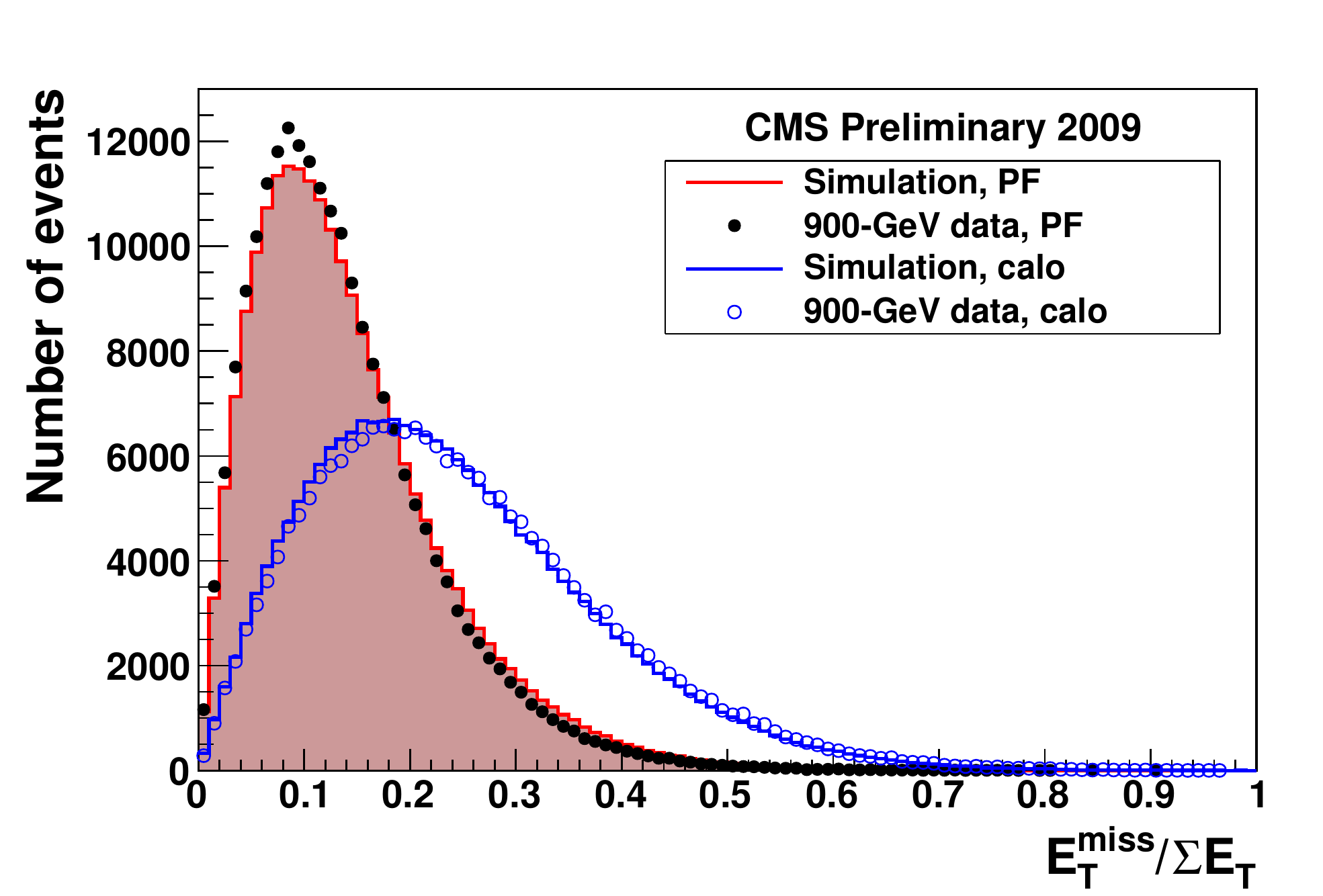}
    \includegraphics[angle=90,width=0.49\textwidth]{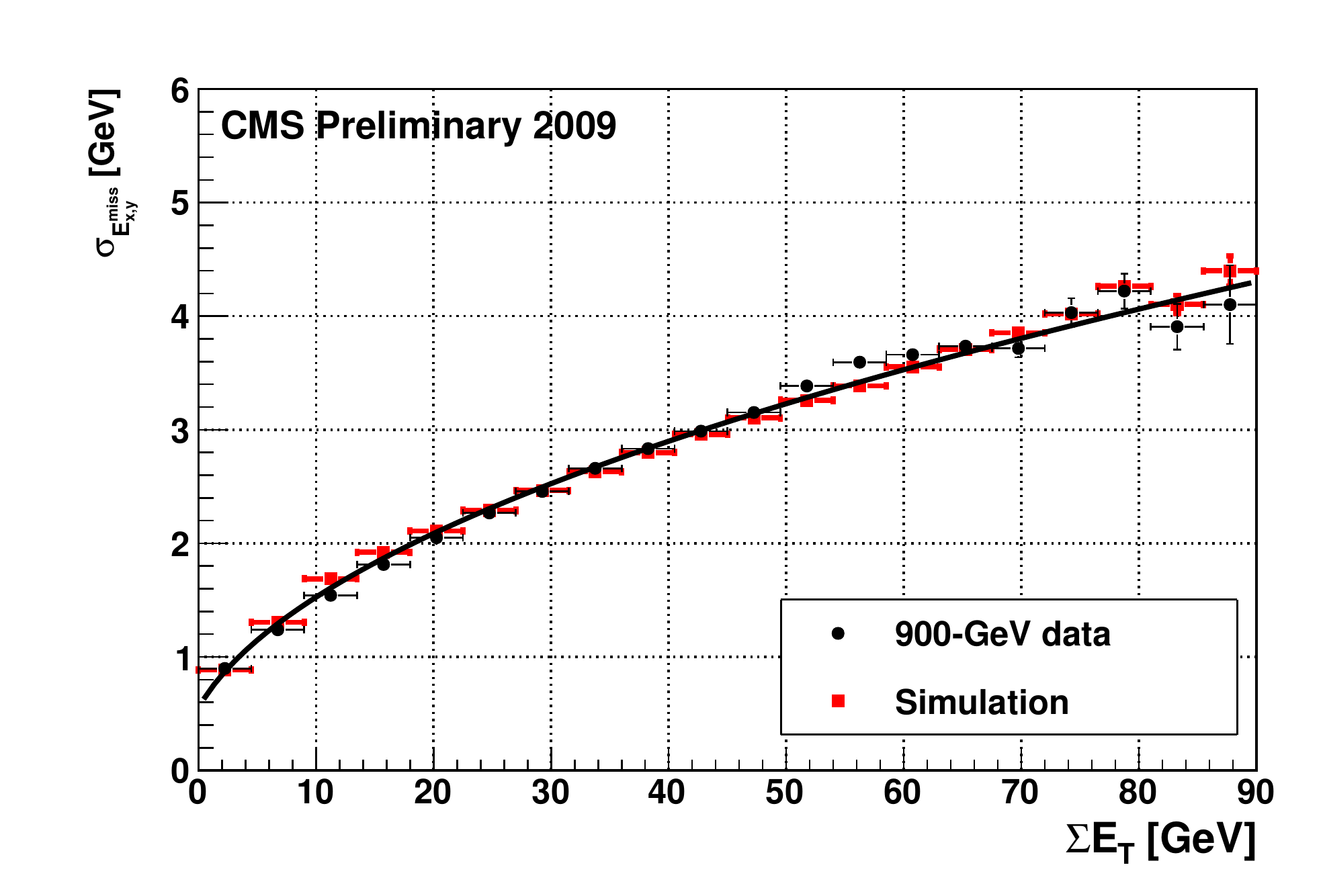}
    \caption{Top: reconstructed MET for caloMET (left), track-corrected MET (center) and particle-flow MET (right). Bottom: MET resolution estimated as MET/$\Sigma E_{\rm T}$ (left), and $E^{\rm miss}_{x,y}/\Sigma E_{\rm T}$ as a function of $\Sigma E_{\rm T}$ (right).}
    \label{fig:METresolution}
  \end{center}
\end{figure}

In all three cases, basic distribution of jet quantities were shown to be
well reproduced by the simulation~\cite{JME-10-001}. Using the same three 
types of input, CMS has started commissioning three algorithms for the
determination of the missing transverse energy (MET), corresponding to 
the modulus of the vector sum of the transverse momenta of all reconstructed
particles in the event~\cite{PFT-10-001,JME-10-002}. 
The distributions of reconstructed MET for all three algorithms are shown
in Fig.~\ref{fig:METresolution} (top row).
Since in this data sample no events are expected with invisible particles
of significant transverse momentum, 
the distribution of MET/$\Sigma E_{\rm T}$ is a measure of the 
resolution of the MET determination, where $\Sigma E_{\rm T}$ is the scalar sum of the transverse
energies of all particles in the event. This distribution, shown in Fig.~\ref{fig:METresolution}(bottom left), gives a good
indication of the improvement in resolution that can be achieved by using the 
Particle-Flow MET, compared to the calorimeter-only MET. Finally the resolution
of the $x$- and $y$-component of MET, $E^{\rm miss}_{x,y}$ is plotted for Particle-Flow MET as a 
function of $\Sigma E_{\rm T}$, and can be parametrized as $\sigma(E^{\rm miss}_{x,y}) = a \oplus b \sqrt{\Sigma E_{\rm T}}$, 
with $a$=0.55~GeV and $b$=45\%, at $\sqrt{s}$= 900~GeV (Fig.~\ref{fig:METresolution}, bottom right).

\section{Transverse-momentum and pseudo-rapidity distributions of charged Hadrons}
\label{sec:dndeta}

\begin{figure}[hbtp]
  \begin{center}
    \includegraphics[width=0.48\textwidth, height=0.5\textwidth]{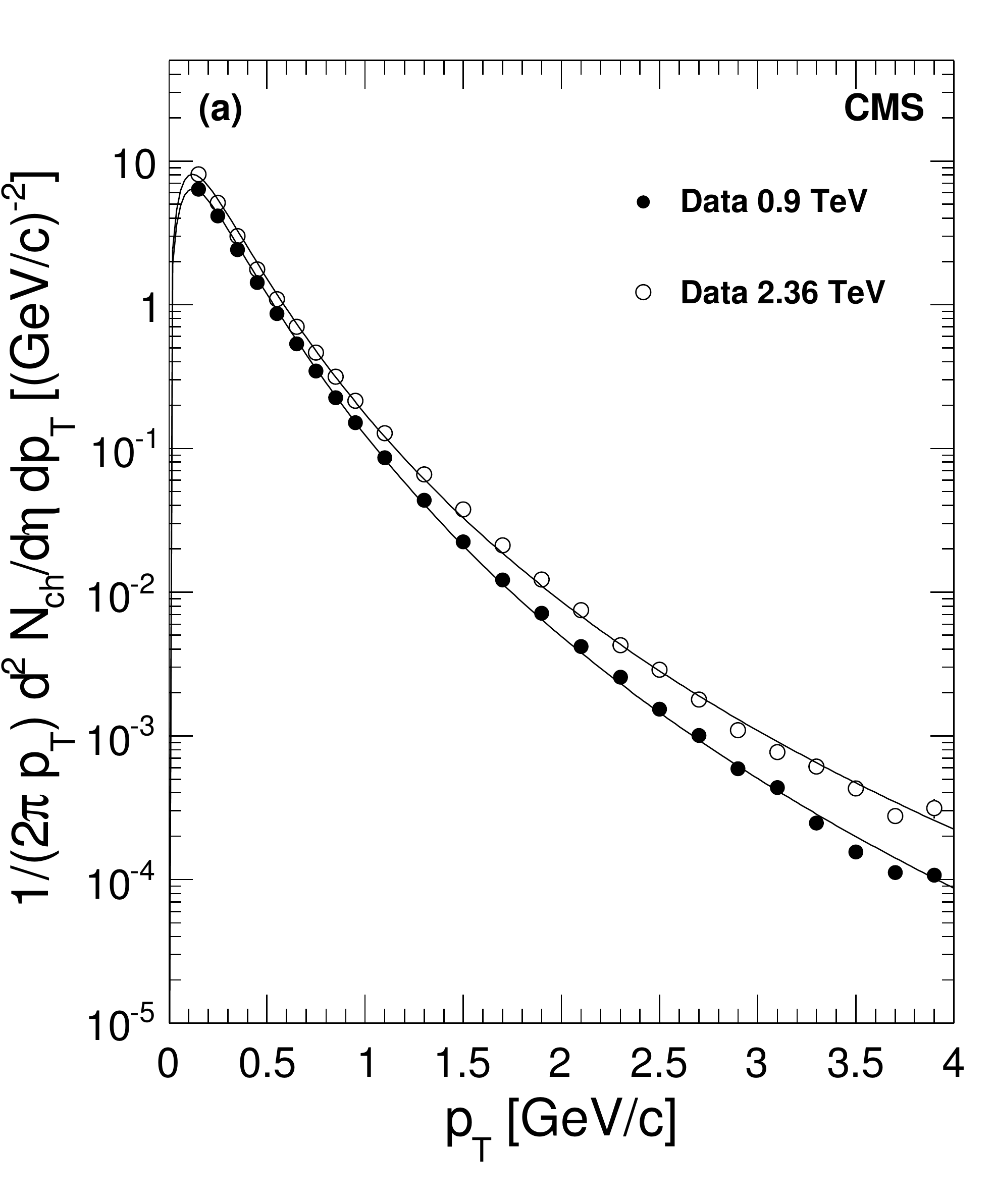}
    \includegraphics[width=0.48\textwidth, height=0.5\textwidth]{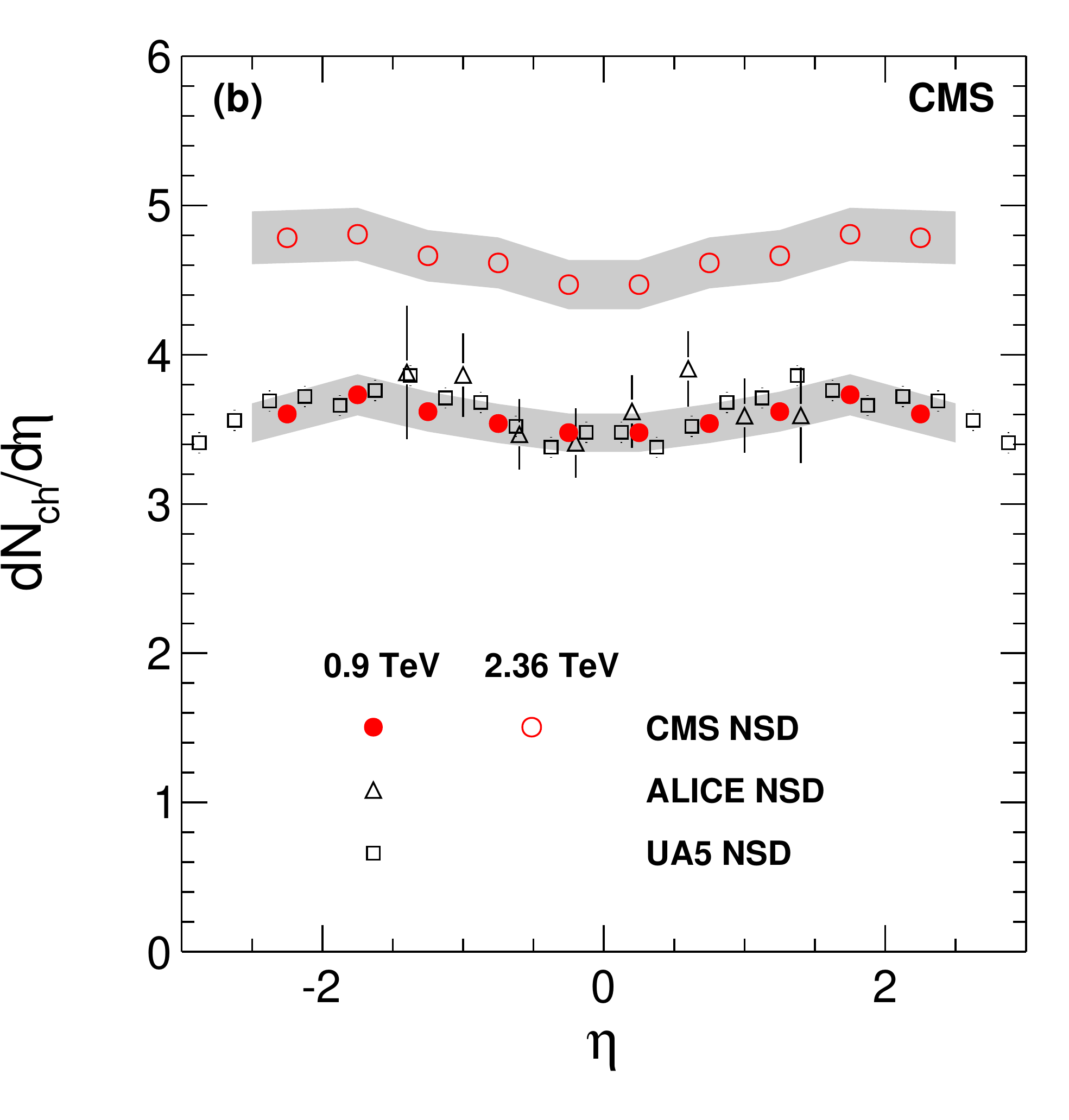}
    \caption{ (a) Measured yield of charged hadrons for $|\eta| < 2.4$ with systematic 
      uncertainties (symbols), fit with an empirical function. 
      (b) Reconstructed pseudo-rapidity density of charged particles averaged over 
      the cluster counting, tracklet and tracking methods (circles), 
      compared to data from the UA5 (open squares) and from the ALICE (open triangles) 
      experiments at 0.9~TeV, and the averaged result over the three methods at 2.36~TeV 
      (open circles). The CMS and UA5 data points are symmetrized in $\eta$. The shaded 
      band represents systematic uncertainties of this measurement, which are 
      largely correlated point-to-point. The error bars on the UA5 and ALICE data 
      points are statistical only.}
    \label{fig:dndeta}
  \end{center}
\end{figure}

The good understanding of the tracker performance allowed a timely publication of
the first physics measurement from collision data performed by CMS: the measurement
of the inclusive charged-hadron transverse-momentum and pseudo-rapidity distributions 
in proton-proton collisions at $\sqrt{s}$ = 900~GeV and 2.36~TeV.\cite{dndeta} For this measurement
three different methods with different sensitivity to potential systematic effects
were combined: pixel cluster counting, pixel tracklets, and 
full track reconstruction. As shown in  Fig.~\ref{fig:dndeta}(a), 
the tracking method allowed reconstruction of tracks down to very low transverse
momentum. The other two methods allow the counting of charged hadrons to even lower values of $p_{\rm T}$.
The combined pseudo-rapidity density result is shown in Fig.~\ref{fig:dndeta}(b).

For non-single-diffractive interactions, the average charged-hadron
transverse momentum was measured to be 0.46$\, \pm \,$0.01 (stat.)$\, \pm \,$0.01
(syst.)~GeV at 0.9~TeV, and 0.50$\, \pm \,$0.01 (stat.)$\, \pm \,$0.01 (syst.) GeV
at 2.36~TeV, for pseudorapidities $| \eta |<$2.4. At these
energies, the measured pseudorapidity densities in the central region,
dN ch/d$\eta$ for $| \eta | <$0.5, are 3.48$\, \pm \,$0.02 (stat.)$\, \pm \,$0.13 (syst.) 
and 4.47$\, \pm \,$0.04 (stat.)$\, \pm \,$0.16 (syst.), respectively. 
The results at 0.9~TeV are in agreement with previous measurements by UA5~\cite{UA5} and 
ALICE~\cite{ALICE},
thus confirming the expectation of near equal hadron production in ${\rm p\bar{p}}$
and pp collisions. The results at 2.36~TeV represent the highest-energy 
measurements at a particle collider to date, at the time of this conference.

\section{Conclusions}
\label{sec:conclusion}

The CMS collaboration has extracted many useful performance results and one 
physics measurement from the first 10~$\mu$b$^{-1}$ of collision data delivered 
by the LHC. Several other physics analyses are in progress. The performance
of the detector at start-up was outstanding. It should however be
noted that the integrated luminosity recorded so far corresponds to less than a millisecond
of data taking at the nominal LHC luminosity, which means that we are still many orders
of magnitude away from a data sample with which CMS can
begin to explore the physics for which the detector was designed. 

Nevertheless, the first results indicate that CMS is in a very good shape to produce high 
quality physics results quickly once more data are recorded in the upcoming 
physics run at a collision energy of 7~TeV.



\section*{Acknowledgments}

I would like to thank the organizers for a very pleasant conference with 
good talks and stimulating discussions. I thank 
my CMS colleagues for preparing the results presented in this report, and 
in particular G.~Tonelli, P.~Janot and A.~De~Roeck for their helpful
suggestions while preparing the talk, as well as N.~Varelas for proofreading
these proceedings. On behalf of CMS I also wish to congratulate our 
colleagues in the CERN accelerator departments for the excellent performance 
of the LHC machine, thank the technical and administrative staff at CERN 
and other CMS institutes, and acknowledge support from: FMSR (Austria); FNRS and FWO (Belgium); CNPq, CAPES, FAPERJ, and FAPESP (Brazil); MES (Bulgaria); CERN; CAS, MoST, and NSFC (China); COLCIENCIAS (Colombia); MSES (Croatia); RPF (Cyprus); Academy of Sciences and NICPB (Estonia); Academy of Finland, ME, and HIP (Finland); CEA and CNRS/IN2P3 (France); BMBF, DFG, and HGF (Germany); GSRT (Greece); OTKA and NKTH (Hungary); DAE and DST (India); IPM (Iran); SFI (Ireland); INFN (Italy); NRF and WCU (Korea); LAS (Lithuania); CINVESTAV, CONACYT, SEP, and UASLP-FAI (Mexico); PAEC (Pakistan); SCSR (Poland); FCT (Portugal); JINR (Armenia, Belarus, Georgia, Ukraine, Uzbekistan); MST and MAE (Russia); MSTDS (Serbia); MICINN and CPAN (Spain); Swiss Funding Agencies (Switzerland); NSC (Taipei); TUBITAK and TAEK (Turkey); STFC (United Kingdom); DOE and NSF (USA).

CMS Physics Analysis Summaries and Detector Performance Summaries are available at\\ http://cdsweb.cern.ch/collection/CMS

\end{document}